\documentclass[5p,sort&compress]{elsarticle}

\usepackage{amsmath}
\usepackage{xspace}

\journal{Physics Letters B}

\newcommand{\beq}{\begin{equation}}
\newcommand{\eeq}{\end{equation}}
\newcommand{\bit}{\begin{itemize}}
\newcommand{\eit}{\end{itemize}}

\def\refeq#1{\mbox{(\ref{#1})}}

\def\reffi#1{\mbox{Fig.~\ref{#1}}}

\def\refta#1{\mbox{Table~\ref{#1}}}

\def\citeres#1{\mbox{Refs.~\cite{#1}}}

\def\ie{i.e.\ }

\newcommand{\rF}{\mathrm{F}}
\newcommand{\rR}{\mathrm{R}}
\newcommand{\rQ}{\mathrm{Q}}
\newcommand{\rT}{\mathrm{T}}
\newcommand{\rd}{\mathrm{d}}
\newcommand{\rs}{\mathrm{s}}

\newcommand{\mur}{\mu_{\rR}}
\newcommand{\muf}{\mu_{\rF}}
\newcommand{\muq}{\mu_{Q}}
\newcommand{\xir}{\xi_{\rR}}
\newcommand{\xif}{\xi_{\rF}}
\newcommand{\xiq}{\xi_{Q}}

\newcommand{\GeV}{\mathrm{GeV}}
\newcommand{\TeV}{\mathrm{TeV}}
\newcommand{\fb}{\mathrm{fb}}
\newcommand{\alphaS}{\alpha_{\rs}}

\def\mathswitchr#1{\relax\ifmmode{\mathrm{#1}}\else$\mathrm{#1}$\fi}
\newcommand{\Pt}{\mathswitchr t}
\newcommand{\Pb}{\mathswitchr b}
\newcommand{\Pc}{\mathswitchr c}
\newcommand{\PW}{\mathswitchr W}
\newcommand{\PH}{\mathswitchr H}
\newcommand{\Pg}{\mathswitchr g}
\newcommand{\Pp}{\mathswitchr p}
\newcommand{\ttbar}{\Pt\bar\Pt}
\newcommand{\bbbar}{\Pb\bar\Pb}
\newcommand{\ccbar}{\Pc\bar\Pc}
\newcommand{\ttbb}{\ttbar\bbbar}

\newcommand{\ttb}{\ttbar\Pb}
\newcommand{\nb}{N_{\Pb}}

\newcommand{\Mt}{m_{\Pt}}
\newcommand{\Mb}{m_{\Pb}}

\newcommand{\LO}{\mathrm{LO}}
\newcommand{\NLO}{\mathrm{NLO}}

\newcommand{\MC}{\mathrm{MC}}
\newcommand{\Sherpa}{{\scshape Sherpa}\xspace}

\newcommand{\Amegic}{{\scshape Amegic++}\xspace}

\newcommand{\OpenLoops}{{\scshape OpenLoops}\xspace}
\newcommand{\SherpaOpenLoops}{{\scshape Sherpa+OpenLoops}\xspace}
\newcommand{\Collier}{{\scshape Collier}\xspace}

\newcommand{\xs}[2]{#1}

\begin{document}


\title{NLO matching for $\ttbb$ production with massive b-quarks}

\author[itpz]{F.~Cascioli}
\ead{cascioli@physik.uzh.ch}

\author[itpz]{P.~Maierh\"ofer}
\ead{philipp@physik.uzh.ch}

\author[itpz]{N.~Moretti}
\ead{moretti@physik.uzh.ch}

\author[itpz]{S.~Pozzorini}
\ead{pozzorin@physik.uzh.ch}

\author[freiburg]{F.~Siegert}
\ead{frank.siegert@cern.ch}

\address[itpz]{Institut f\"ur Theoretische Physik, Universit\"at Z\"urich, 8057 Z\"urich, Switzerland}

\address[freiburg]{Physikalisches Institut, Albert-Ludwigs-Universit{\"a}t Freiburg, D-79104 Freiburg, Germany}


\begin{abstract}
Theoretical uncertainties in the simulation of $\ttbb$
production represent one of the main obstacles that still hamper the
observation of Higgs-boson production in association with top-quark pairs in
the $\PH\to \bbbar$ channel.  In this letter we present a next-to-leading
order (NLO) simulation of $\ttbb$ production with massive b-quarks matched to the
\Sherpa parton shower.  This allows one to extend NLO predictions to
arbitrary $\ttbb$ kinematics, including the case where one or both b-jets
arise from collinear $\Pg\to\bbbar$ splittings.  We find that this splitting
mechanism plays an important role for the $\ttbar\PH(\bbbar)$
analysis.
\end{abstract}


\maketitle

\sloppy

The recent discovery of the Higgs boson and first measurements of its
interactions permit to probe the mechanism of spontaneous symmetry breaking, 
by which elementary particles acquire their mass~\cite{Aad:2012tfa,Chatrchyan:2012ufa}.
Data collected in the first run of the LHC provide significant sensitivity to Higgs-boson
interactions with force carriers---gluons, photons, Z and W bosons---while 
constraints on Higgs-couplings to matter particles---leptons and
quarks---are less stringent and mostly stemming from indirect effects on
Higgs--gluon and Higgs--photon couplings.  
The direct investigation of Higgs-boson couplings to quarks and leptons
will thus represent a crucial further step 
towards a complete understanding of the origin of
mass.
In
this context, the reaction $\Pp\Pp\to \Pt\bar \Pt \PH(\Pb\bar
\Pb)$, i.e.~Higgs-boson
production in association with a top-quark pair with subsequent Higgs-boson
decay into a bottom-quark pair, provides a unique
opportunity to test the mass-generation mechanism in the heavy-quark sector.
This process is notoriously very challenging due to the presence of four
b-quarks in the final state, which hampers a correct identification of the
Higgs-boson mass peak.  As a result, the $\ttbar \PH$ signal is strongly
contaminated by background contributions from top-quark pair production in
association with light-, charm- and bottom-jet  pairs.  The large
uncertainty in the Monte-Carlo simulations of these multi-particle QCD
backgrounds represents one of the main bottlenecks of the present $\ttbar\PH(\bbbar)$
analyses~\cite{Chatrchyan:2013yea, ATLAS:2012cpa}, and the availability of state-of-the art 
theory predictions for $\ttbar jj$, $\ttbar\ccbar$, and $\ttbar\bbbar$ production is a key
prerequisite to improve the sensitivity to the $\ttbar\PH(\bbbar)$ signal. 
In the case of the irreducible $\ttbb$ background, theory predictions play
an especially important role, since the lack of sufficiently distinctive
kinematic features and the rather small cross section do not allow for an
efficient $\ttbb$ measurement in a signal-free control region.

\begin{figure}
  \begin{center}
    \includegraphics{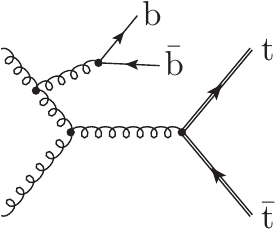}
    \hspace{5mm}
    \raisebox{-2.8mm}{\includegraphics{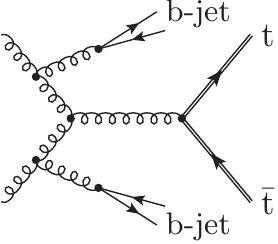}}
  \end{center}
  \vspace{-2mm}
  \caption{Tree topologies corresponding to $\ttbb$ production via single hard (left) or double collinear (right) $\Pg\to\bbbar$ splitting.}
  \label{fig:topologies} 
\end{figure}

NLO calculations for $\ttbb$
\cite{Bredenstein:2009aj,Bredenstein:2010rs,Bevilacqua:2009zn,Worek:2011rd} and $\ttbar jj$
\cite{Bevilacqua:2010ve,Bevilacqua:2011aa} production can
reduce perturbative uncertainties from 70--80\% down to 15--20\%.  However, in
order to be applicable to the experimental analyses, these calculations need
to be matched to parton showers.  
Matched NLO predictions for $\Pp\Pp\to\ttbar+\le 1$ jets, 
with consistent merging of 0- and 1-jet final states,
have been presented in~\cite{Hoeche:2013mua}, and
first technical results towards NLO
matched $\ttbb$ production have been discussed in \cite{Kardos:2013vxa},
where the NLO calculation of \cite{Bevilacqua:2009zn} was matched at the
level of the first shower emission with the {PowHeg}
approach~\cite{Frixione:2007vw}.
In this letter, we present a fully-showered NLO simulation of 
$\ttbb$ production. Besides matching NLO matrix elements to the
parton shower with the MC@NLO method \cite{Frixione:2002ik},
for the first time we also include finite b-quark mass effects.
This represents the first complete NLO-matched  
simulation with four (massive) coloured particles in the final state.
Using massive b-quarks we can extend the simulation 
to the whole $\ttbb$ phase space, thereby including also
$\ttbar+1\,\Pb$-jet contributions with an unresolved (soft or collinear)
b-quark, which play an important role in the $\ttbar\PH(\bbbar)$ analysis.
Moreover, matching massive NLO matrix elements 
to the parton shower gives access to novel $\ttbar+\Pb$-jets production mechanisms, 
where b-jets arise from hard gluons via collinear $\Pg\to\bbbar$ splittings.
In particular, one can describe $\ttbar+2\,\Pb$-jet
events where both b-jets  originate from $\Pg\to\bbbar$ splittings (see Fig.~\ref{fig:topologies}).
For this kind of configurations---which turn out to be quite important---the finite 
b-quark mass allows one to obtain an NLO accurate description of 
the first $\Pg\to\bbbar$ splitting, while simulations with massless b-quarks 
must rely on $\ttbar\Pg\Pg$ matrix elements plus
pure parton-shower splittings in the collinear regions.

The presented simulation has been prepared within the \SherpaOpenLoops
framework~\cite{Gleisberg:2008ta,Cascioli:2011va,Cascioli:2013gfa}, which 
supports the fully automated simulation of any Standard-Model process at
NLO QCD, including matching to the parton shower and multi-jet merging.
The \OpenLoops~\cite{Cascioli:2011va} program is a one-loop generator based on a
novel numerical recursion, which is formulated in terms of loop-momentum
polynomials called ``open loops'' and allows for a fast evaluation of
scattering amplitudes with many external particles.\footnote{A public implementation of {\OpenLoops} will appear in the next future~\cite{openloops}.}  It uses the \Collier
library~\cite{collier} for the numerically stable evaluation of tensor
integrals~\cite{Denner:2002ii,Denner:2005nn} and scalar
integrals~\cite{Denner:2010tr}.  Real-emission contributions,
infrared subtractions based on the Catani--Seymour (CS) 
technique~\cite{Catani:1996vz,Catani:2002hc}, and
phase-space integration are handled by \Sherpa~\cite{Gleisberg:2008ta} and
\Amegic~\cite{Krauss:2001iv}.
The NLO corrections are matched to the \Sherpa parton
shower~\cite{Schumann:2007mg} using the \Sherpa
formulation~\cite{Hoeche:2011fd,Hoeche:2012fm} of the MC@NLO
method~\cite{Frixione:2002ik}.\footnote{In the following, MC@NLO always
refers to the algorithm of~\citeres{Hoeche:2011fd,Hoeche:2012fm} and its
implementation within \Sherpa.}  The essence of the MC@NLO approach is
encoded in the following formula for the no-emission and
first-emission contributions to the expectation value of a generic
observable~\cite{Hoeche:2012fm}, 
\begin{eqnarray}
\lefteqn{\langle \mathcal{O} \rangle = \int \rd \Phi_B
\biggl[B(\Phi_B)+V(\Phi_B)+I(\Phi_B)\biggr] U(t_0,\muq^2)
}\quad\nonumber\\&&{} \hspace{-17mm}+ \int\rd \Phi_R\biggl[
 R(\Phi_R)-
\sum_{ijk} D_{ijk}(\Phi_R)\theta(\muq^2-t)
\biggr]
\mathcal{O}(\Phi_R).
\label{eq:mcatnlo}%
\end{eqnarray}
The terms $B(\Phi_B)$ and $V(\Phi_B)$
represent Born and virtual matrix-element
contributions to the Born phase space $\Phi_B$, while
$R(\Phi_R)$ denotes real-emission matrix-element
contributions to the corresponding phase space $\Phi_R$.
Similarly as for NLO calculations, 
infrared singularities are removed
from the $\Phi_R$ phase space via local subtraction terms
$D_{ijk}(\Phi_R)$ and added back
to the virtual contributions
in the form
\begin{eqnarray}
I(\Phi_B)&=&\sum_{ijk}\int\rd \Phi_{R|B} D_{ijk}(\Phi_{R})\theta(\muq^2-t),
\label{eq:Ioper}
\end{eqnarray}
where each subtraction term is integrated over 
a factorised phase space $\Phi_{R|B}$ 
associated with a $\Phi_R\to\Phi_B$ mapping.
In fixed-order calculations, to achieve an exact cancellation of the
subtraction terms, events associated with 
$D_{ijk}(\Phi_R)$ must be attributed to the Born phase space
according to the appropriate $\Phi_R\to\Phi_B$ mapping.
In contrast, in the MC@NLO approach
$D_{ijk}(\Phi_R)$ contributions are handled as
genuine real-emission events, and the
resulting mismatch of the form $D_{ijk}(\Phi_R)\left[\mathcal{O}(\Phi_R)-
\mathcal{O}(\Phi_B)\right]$
is compensated, to order $\alphaS$,
by $\Phi_B\to \Phi_R$ migrations that result from 
parton-shower emissions.
The first shower emission is described by
\begin{eqnarray}
\lefteqn{
U(t_0,\muq^2)
=
\Delta(t_0,\muq^2)
\mathcal{O}(\Phi_B)
}\quad&&\nonumber
\\&&\hspace{-8mm}{}
+
\sum_{ijk}\int_{t_0}^{\muq^2}
\rd \Phi_{R|B}
\frac{D_{ijk}(\Phi_R)}{B(\Phi_B)}
\Delta(t,\muq^2)
\mathcal{O}(\Phi_R),
\label{eq:css}
\end{eqnarray}
where the second line corresponds to the first-emission 
probability, and the Sudakov form factor $\Delta(t_0,\muq^2)$ 
represents its no-emission counterpart.
The parton shower is driven by the 
evolution variable $t$. It starts at the resummation scale
$\muq^2$ and stops when $t$ reaches the infrared cut-off
$t_0$. 
The key principle, by means of which the MC@NLO approach
preserves NLO accuracy up to
the first emission, is the correspondence between the splitting kernels of
the parton shower and the terms $D_{ijk}$ that are subtracted from the real
emission.  In \Sherpa this is achieved by using CS dipoles $D_{ijk}$ both
as subtraction terms and as splitting kernels of the parton shower.  More
precisely, the kernels of the shower are given by the spin-averaged CS
dipoles, taken in the large-$N_\mathrm{c}$ limit.  In addition, to obtain a
fully consistent matching, the first shower emission is supplemented by
exact spin and colour correlations~\cite{Hoeche:2011fd}.  The MC@NLO
matching can be regarded as an effective subtraction of the first shower
emission, and, similarly as for the shower, also the subtraction terms in
\refeq{eq:mcatnlo} and \refeq{eq:Ioper} must be restricted to the kinematic
region $t<\muq^2$.  Finally, no-emission and first-emission events generated
according to \refeq{eq:mcatnlo}--\refeq{eq:css} are used as seeds for
subsequent shower emissions.

\begin{table*}[t]
  \vspace*{0.3ex}
  \begin{center}
    \begin{tabular}{ll@{\hspace{12mm}}l@{\hspace{12mm}}l}
~&  ttb 
&  ttbb
&  ttbb$({m_{\Pb\Pb}>100)}$
\\[.5mm] \hline
{{$\sigma_\LO[\fb]$}} \phantom{\huge{X}}
&  $\xs{2644}{(1)}^{+71\%}_{-38\%}$$^{+14\%}_{-11\%}$                 
&  $\xs{463.3}{(2)}^{+66\%}_{-36\%}$$^{+15\%}_{-12\%}$                 
&  $\xs{123.4}{(1)}^{+63\%}_{-35\%}$$^{+17\%}_{-13\%}$                 
\\[1mm] \hline
{$\sigma_\NLO[\fb]$} \phantom{\huge{X}} 	
&  $\xs{3296}{(5)}^{+34\%}_{-25\%}$$^{+5.6\%}_{-4.2\%}$                 
&  $\xs{560}{(1)}^{+29\%}_{-24\%}$$^{+5.4\%}_{-4.8\%}$                 
&  $\xs{141.8}{(0.4)}^{+26\%}_{-22\%}$$^{+6.5\%}_{-4.6\%}$                 
\\ 
{$\sigma_\NLO/\sigma_\LO$} \phantom{\huge{X}}
&  $1.25$                                  
&  $1.21$                                  
&  $1.15$                                  
\\[.5mm] \hline
{$\sigma_\MC[\fb]$} \phantom{\huge{X}}
&  $\xs{3313}{(3)}^{+32\%}_{-25\%}$$^{+3.9\%}_{-2.9\%}$   
&  $\xs{600}{(1)}^{+24\%}_{-22\%}$$^{+2.0\%}_{-2.1\%}$   
&  $\xs{181.0}{(8)}^{+20\%}_{-20\%}$$^{+8.1\%}_{-6.0\%}$   
\\ 
{$\sigma_\MC/\sigma_\NLO$} \phantom{\huge{X}} 	
&  $1.01$                                  
&  $1.07$                                  
&  $1.28$                                  
\\[.5mm] \hline 
{$\sigma^{2\Pb}_\MC[\fb]$}  \phantom{\huge{X}}
&  $\xs{3299}{(5)}$   
&  $\xs{552}{(1)}$   
&  $\xs{146}{(1)}$   
\\ 
{$\sigma^{2\Pb}_\MC/\sigma_\NLO$} \phantom{\huge{X}}
&  $1.00$                                  
&  $0.99$                                
&  $1.03$                                
\\[.5mm]\hline
    \end{tabular}
  \end{center}
  \caption{Cross sections with standard $\Pt\Pt\Pb$ and $\Pt\Pt\Pb\Pb$ cuts and with an additional cut, $m_{\Pb\Pb}>100~\GeV$. Full MC@NLO predictions ($\sigma_\MC$) are compared to results obtained with parton-shower $\Pg\to\bbbar$ splittings switched off ($\sigma^{2\Pb}_{\MC}$). The first and second uncertainty represent $\xi_\rR$ and $\xi_\rF$ variations. In the MC@NLO case, the latter is combined with $\xi_\rQ$ variations in quadrature.}
  \label{tab:XS}
\end{table*}

In the following, we present and compare LO, NLO and MC@NLO 
simulations of  $\ttbb$ production at the $8~\TeV$ LHC. The
results are based on a \Sherpa 2.0 pre-release
version.\footnote{This version corresponds to SVN revision 23546, which 
implements a recent tune of the \Sherpa parton shower to LEP data.}
Hadronisation and underlying events are not considered,
and top quarks are treated as stable particles with mass
$\Mt=173.2~\GeV$. While spin-correlated $\Pt\to\PW\Pb$ decays
can be simulated in a fully automated way, 
omitting top decays permits us to focus on the
behaviour of those b-jets that arise from QCD interactions, 
and that involve many more subtleties from the 
viewpoint of the theoretical simulation and its uncertainties.
Consistently with the use of a finite b-quark mass, $\Mb=4.75~\GeV$,
we employ four-flavour parton distributions. Specifically,
at NLO\,(LO) QCD the LHApdf implementation of the MSTW2008NLO\,(LO) parton
distributions~\cite{Martin:2009iq} and the corresponding $\alphaS$ values
are used.
While the four-flavour running of $\alphaS$ misses top- and bottom-quark loop effects,
corresponding $\mathcal{O}(\alphaS)$ contributions are consistently included 
in the virtual corrections via zero-momentum subtraction of 
the heavy-quark loops in the renormalisation of $\alphaS$.

As renormalisation scale we employ
the geometric average of the top-quark and 
b-quark transverse energies,\footnote{
Note that a dynamical QCD scale defined in terms of b-quark momenta 
is infrared safe for $\Mb>0$, while for massless b-quarks 
a scale based on b-jet momenta should be used.
}
\beq
\mur^4 = \xir^4 \prod_{i=\Pt,\bar\Pt,\Pb,\bar\Pb} E_{\rT,i}
= \xir^4\prod_{i=\Pt,\bar\Pt,\Pb,\bar\Pb} \sqrt{m^2_i + p^2_{\rT,i}}\;,
\label{eq:muR}
\eeq
which represents a natural generalisation of the dynamical scale 
$\mu^2=\Mt\sqrt{p_{\rT,\Pb}p_{\rT,\bar\Pb}}$ introduced in~\cite{Bredenstein:2010rs}.
The default scale corresponds to $\xir=1$, and $\xir$ parametrises
scale variations.
To NLO accuracy, this choice corresponds to 
$\alphaS^4(\mur)\simeq\prod_i \alphaS(E_{\rT,i})$ and guarantees that the 
strong-coupling factors associated to the production of the various 
final-state objects adapt to the respective transverse energies.
The factorisation and resummation scales, which define the 
available phase space for QCD radiation, are related to the
average top-quark transverse energy via
\beq
\muf =\frac{\xif}{2}(E_{\rT,\Pt}+E_{\rT,\bar\Pt}),\qquad
\muq = \xiq \muf. 
\eeq
The default scale choice corresponds to $\xif=\xiq=1$, 
and $\xif$ parametrises correlated variations of $\muf$ and $\muq$, while
$\xiq$ controls additional variations of $\muq$ with fixed 
$\muf$.
QCD partons, including b-quarks and excluding only top-quarks, are recombined
into IR-safe jets using the anti-$k_\rT$ algorithm~\cite{Cacciari:2008gp} 
with jet-resolution
parameter $R=0.4$.  Events are categorised according to the number $\nb$
of reconstructed b-jets with $p_\rT>25~\GeV$ and $|\eta_\Pb|<2.5$.  In this
respect, we classify as b-jet any jet involving at least a b-quark, which
includes also the case of collimated $\bbbar$ pairs resulting from the
splitting of energetic gluons.  This is, at least experimentally, the most
realistic b-jet definition, and its implementation at NLO is possible only in
presence of massive b-quarks.  In fact, in calculations with massless
b-quarks, collimated $\bbbar$ pairs must be handled as gluon-jets in
order to avoid collinear singularities.

To investigate NLO and MC@NLO correction effects we
considered an exclusive ttbb sample, with events involving $\nb\ge 2$ b-jets,
and a more inclusive ttb sample with $\nb\ge 1$.  
For the ttbb sample an additional analysis is performed  
with a cut on the invariant mass of the first and second
b-jet, $m_{\Pb\Pb}>100~\GeV$, which corresponds to the
$\ttbar\PH(\bbbar)$ signal region.
The respective LO, NLO and
MC@NLO cross sections are reported in \refta{tab:XS}.  In
order to isolate contributions arising from b-quarks emitted by the
parton shower, we also present MC@NLO predictions generated in absence of
$\Pg\to\bbbar$ parton-shower splittings.  
Scale uncertainties are assessed 
via independent factor-two variations
of $\xir$ and $\xif$. Additional scale uncertainties related to the 
parton shower are included via $\xiq=2^{\pm 1/2}$ variations 
of the resummation scale
and are combined in quadrature 
with $\xif$ variations.

\begin{figure*}[t]
  \begin{center}
    \includegraphics[width=.48\textwidth]{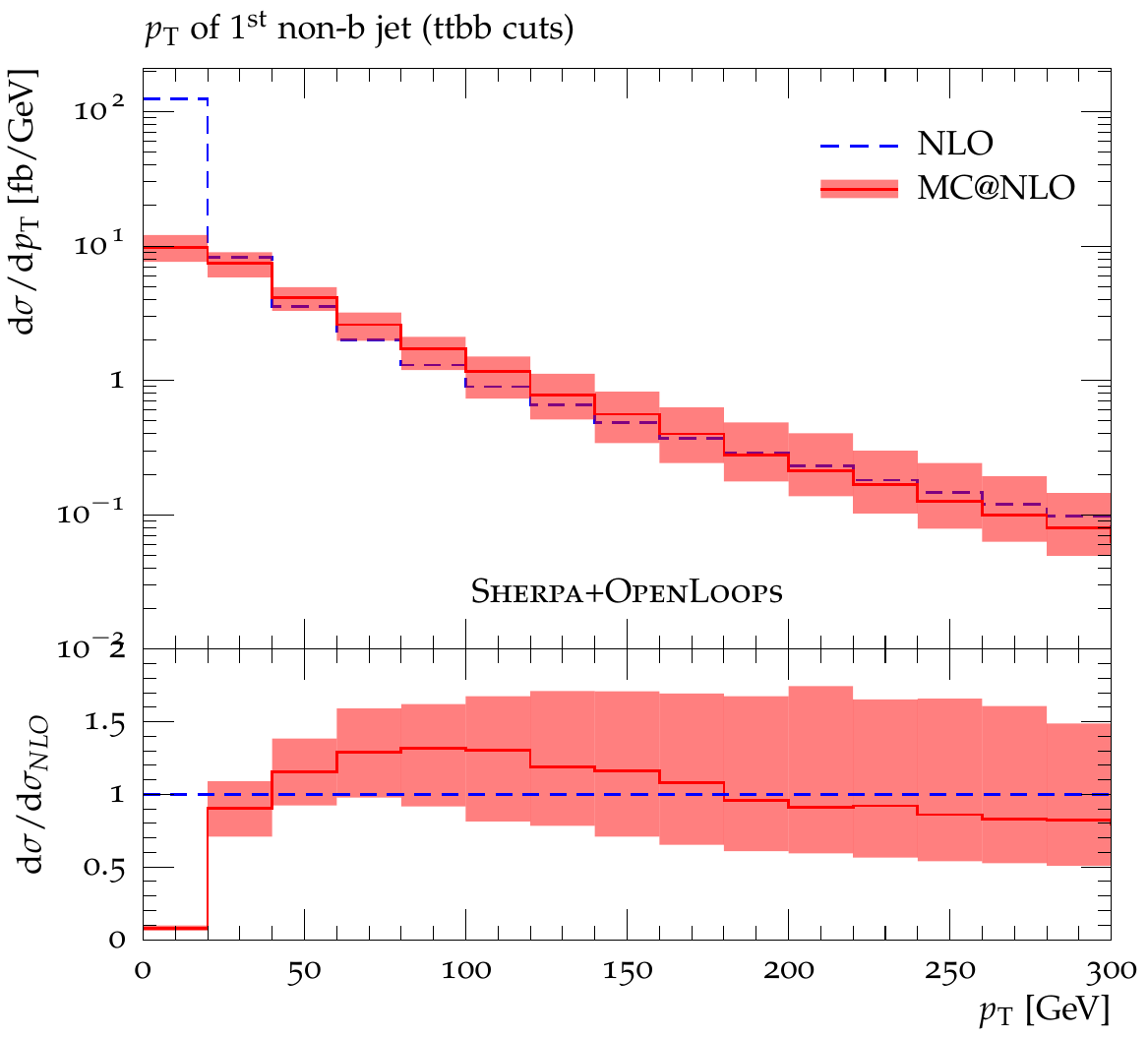}
    \includegraphics[width=.48\textwidth]{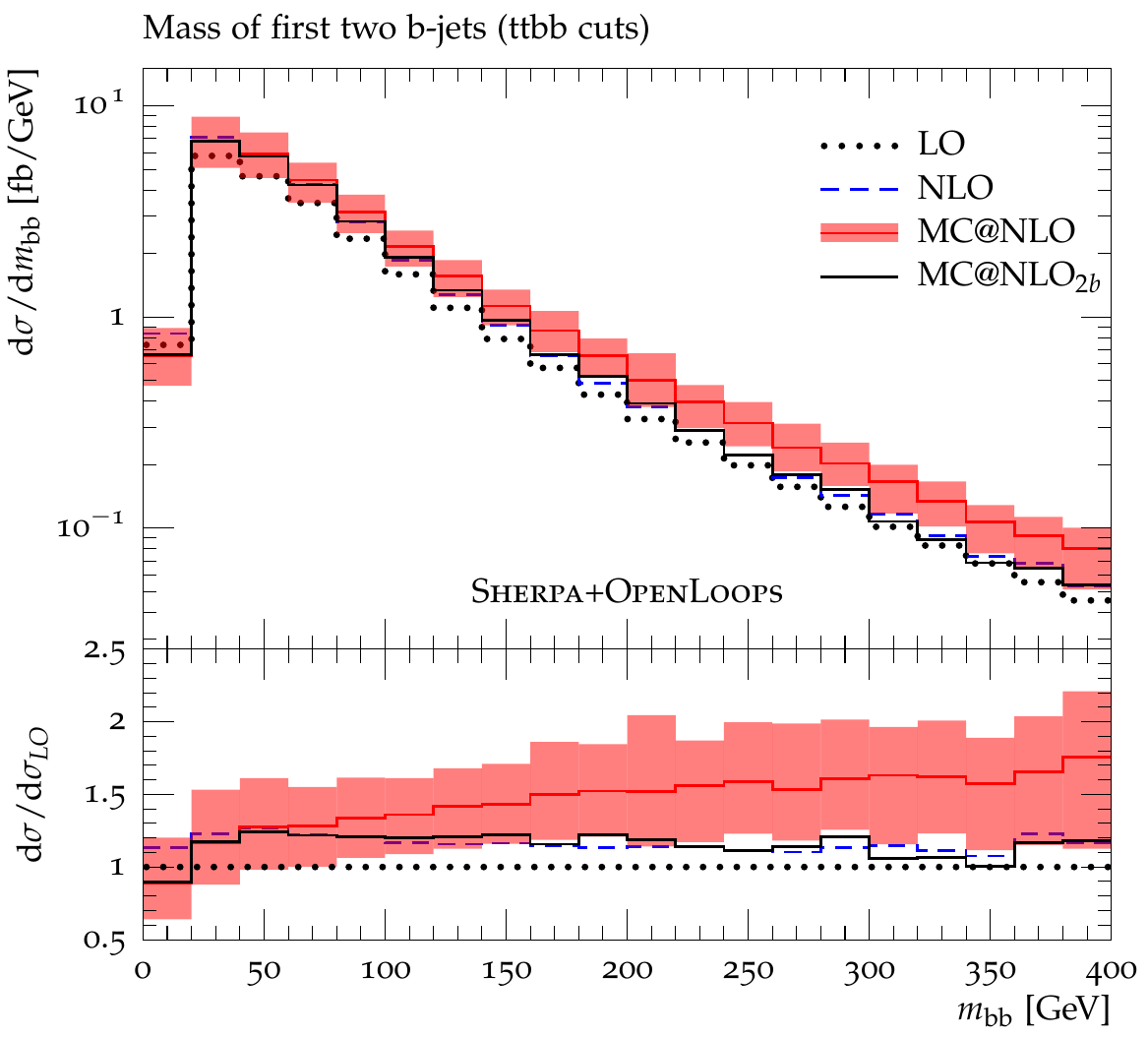}
  \end{center}
  \caption{Transverse momentum of the first light jet and invariant mass of the first two b-jets with standard ttbb cuts.
{The MC@NLO bands display the combination
in quadrature of $\mur$, $\muf$ and $\muq$ scale variations. The
MC@NLO$_{2b}$ curve is obtained by switching off
$\Pg\to\bbbar$ splittings in the parton shower.}
}
  \label{fig:ttbb}
\end{figure*}

Fixed-order results in~\refta{tab:XS} feature 
NLO $K$-factors close to $1.2$,
with $\pm 0.05$ variations depending on the selection cuts.  
This is consistent with the 
$\mathcal{O}(20\%)$ contribution of b-quarks to the running of
$\alphaS^4(\mu)$ from $m_\Pb$ to $\mu_\rR$,
and with the fact that the corresponding $K$-factor
in the five-flavour scheme, where b-quark 
contributions are included in the running of $\alphaS$, 
is very close to one~\cite{Heinemeyer:2013tqa}.
In this respect, let us note that a fully consistent resummation of $\ln(\mu_\rR/\Mb)$
terms associated with the running of $\alphaS$ would 
increase the $\ttbb$ NLO cross section by about
$9\%$ as compared to standard 4F-scheme predictions presented in this letter.
This estimate was obtained using a modified set of MSTW four-flavour
PDFs with five active flavours in the evolution of $\alphaS$.

Scale uncertainties 
in~\refta{tab:XS}
are dominated by renormalisation-scale variations and
decrease from about 60--70\% at LO to 20--30\% at NLO.
Scale variations at NLO and MC@NLO level are rather similar.
In presence of 
standard ttb and ttbb cuts,
matching to the parton shower 
shifts the NLO cross section by 
only $1\%$ and $6\%$, respectively. However,
the MC@NLO correction to $\ttbb$ finals states is quite 
sensitive to the invariant mass of the ${\bbbar}$ pair and 
turns out to be enhanced by a factor four 
in the region $m_{\bbbar}>100\GeV$, which is relevant for Higgs-boson searches.
This MC@NLO effect---which clearly 
exceeds the magnitude of the Higgs signal in the present 
$\ttbar\PH(\bbbar)$ analyses~\cite{Chatrchyan:2013yea, ATLAS:2012cpa}---tends to 
disappear if $\Pg\to\bbbar$ splittings are switched off in the parton shower.\footnote{
Note that only full MC@NLO predictions should be regarded as  
physical, while results without $\Pg\to\bbbar$ parton-shower splittings 
are showed only for technical aims, namely to illustrate the relevance of multiple $\bbbar$ production.
}
As discussed below, various features indicate that this effect is 
dominated by the double-splitting mechanism depicted in \reffi{fig:topologies}.b.

\begin{figure*}[t]
  \begin{center}
    \includegraphics[width=.48\textwidth]{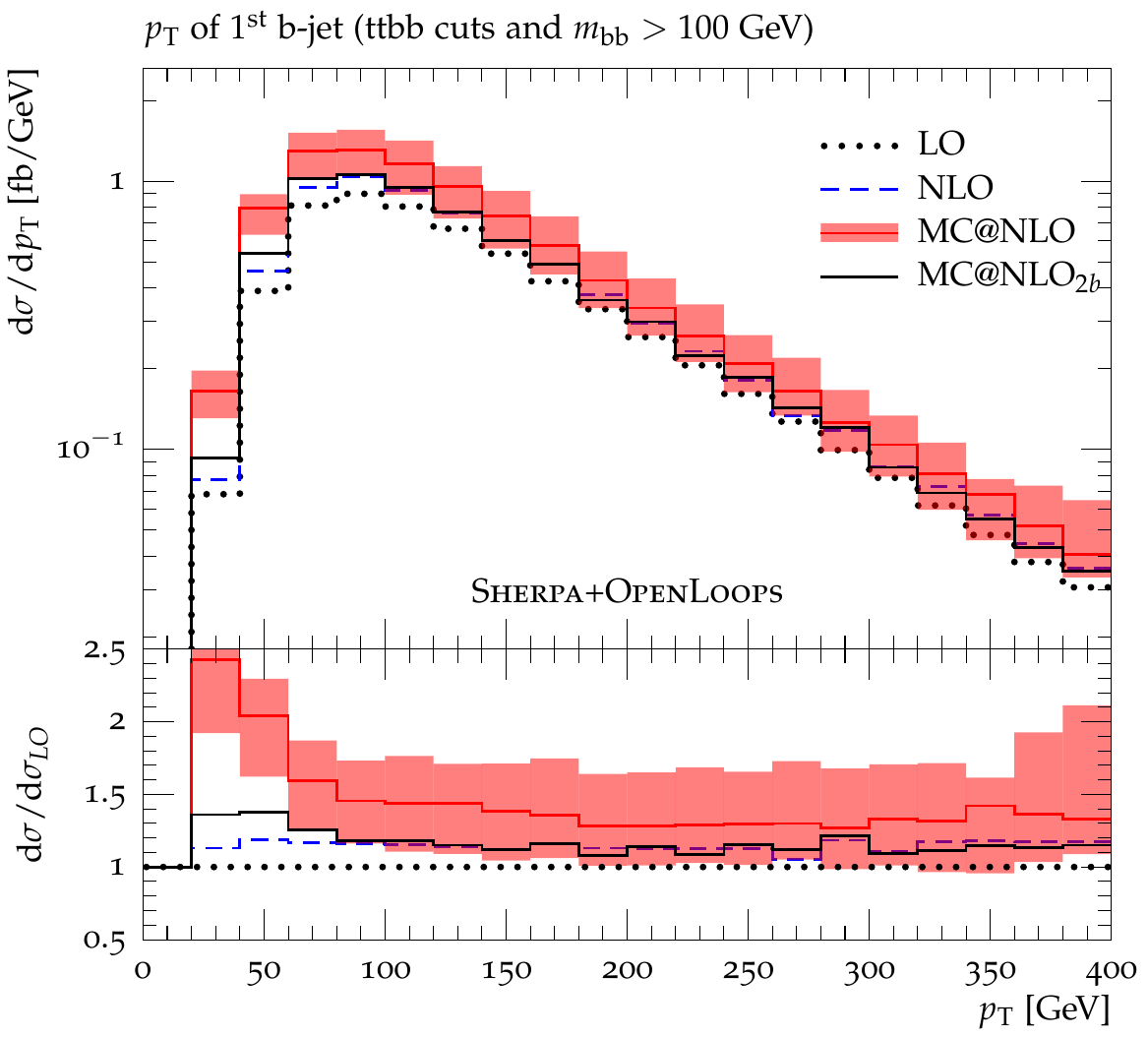}
    \includegraphics[width=.48\textwidth]{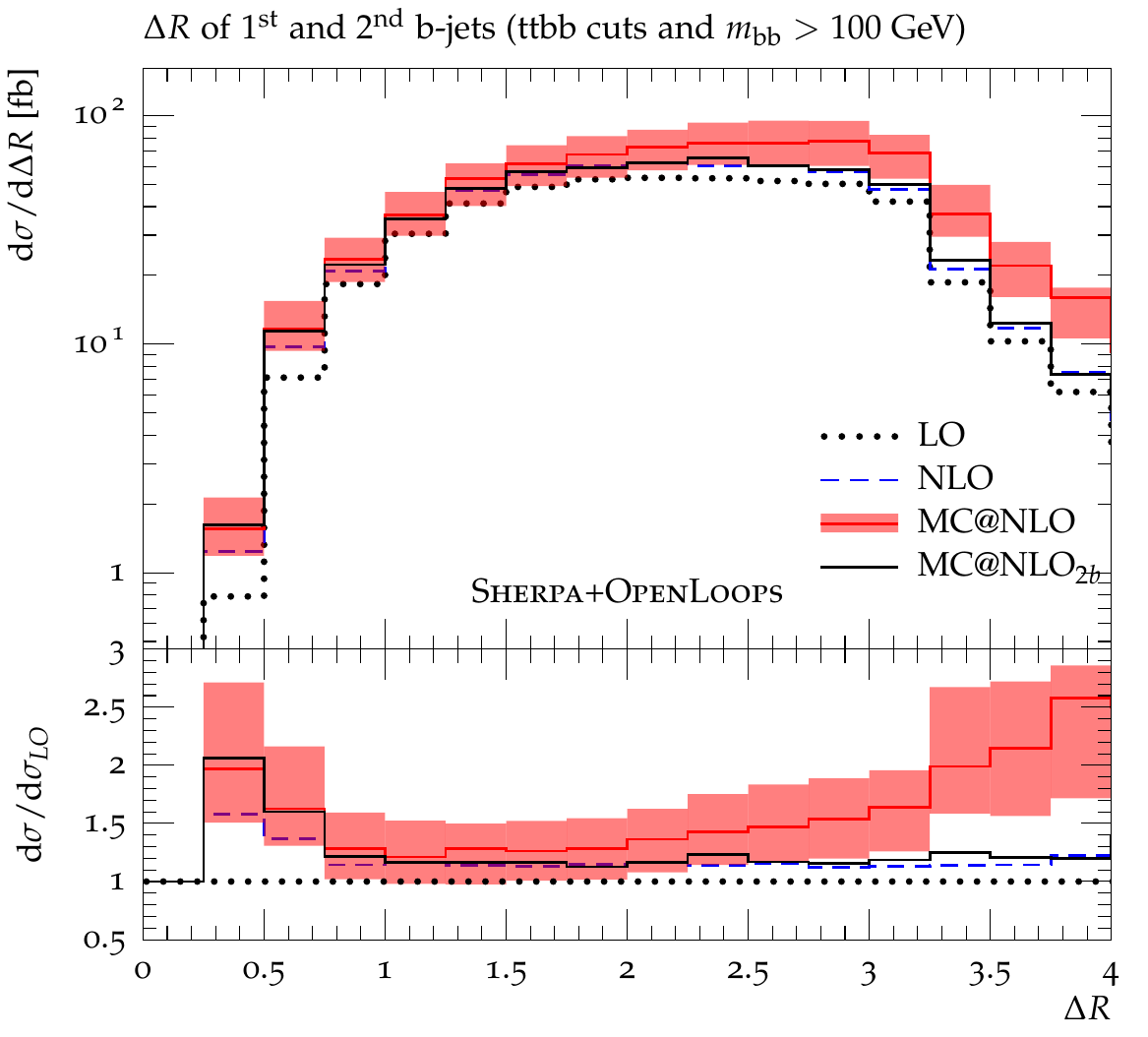}
  \end{center}
  \caption{Transverse momentum of the first b-jet and $\Delta R$ separation of the first two b-jets with standard ttbb cuts and $M_{\Pb\Pb}>100~\GeV$.
{The MC@NLO bands display the combination
in quadrature of $\mur$, $\muf$ and $\muq$ scale variations. The 
MC@NLO$_{2b}$ curve is obtained by switching off
$\Pg\to\bbbar$ splittings in the parton shower.}
}
  \label{fig:ttbb100}
\end{figure*}

The differential distributions in Figs.~\ref{fig:ttbb} and \ref{fig:ttbb100} 
provide examples of nontrivial matching corrections.
Standard ttbb cuts are applied, and the MC@NLO bands display the combination
in quadrature of $\mur$, $\muf$ and $\muq$ scale variations.  
The corresponding uncertainties are typically around 30\% and tend to increase
in the tails, also due to statistical fluctuations.
The transverse
momentum of the first non-b jet (Fig.~\ref{fig:ttbb}.a) shows the 
typical MC@NLO behaviour. At transverse momenta above the resummation
scale, where the parton shower stops emitting, MC@NLO and NLO 
predictions agree well. The fixed-order infrared singularity at small $p_\rT$ is
consistently damped by the Sudakov form factor, and
Sudakov effects start to be important already at $p_\rT\sim 50~\GeV$.  This
reflects the presence of intense QCD radiation resulting from the
gluon-gluon initial state and from the high center-of-mass energy of the
$\ttbb$ system.
In the intermediate $p_\rT$ region we observe an MC@NLO
correction of about $+30\%$ wrt.~NLO. This can be attributed 
to $\Pg\to\bbbar$ parton-shower splittings and to the 
enhancement of the
first shower emission that results from the $(B+V+I)$ term in
\refeq{eq:mcatnlo}.
The precise position and magnitude of the MC@NLO/NLO maximum depend on the
choice of the renormalisation and resummation scales, and scale
variations permit assessing related higher-order uncertainties inherent in the
matching procedure.

Figure~\ref{fig:ttbb}.b confirms that matching corrections
are quite sensitive to the 
invariant mass of the first two b-jets.
The MC@NLO/NLO ratio grows with $m_{\Pb\Pb}$ and reaches 
25--30\% in the Higgs-signal region, $m_{\Pb\Pb}\sim 125~\GeV$.
This enhancement at high invariant mass can be 
attributed to $\ttbar+2\,\mathrm{b}$-jets
production via double $\Pg\to\bbbar$ splittings, 
since this mechanism is kinematically favoured
by the fact that the probability that two hard gluons 
split into collinear $\bbbar$ pairs does not decrease 
when the invariant mass of the gluon pair grows.
This interpretation is {confirmed}  by the fact that
the shape of the MC@NLO $m_{\Pb\Pb}$ distribution becomes 
 {almost identical} to the NLO one if $\Pg\to\bbbar$ splittings are switched off
in the parton shower. 
Further evidence of the correctness of the above picture
is provided by the fact that the MC@NLO
excess increases with the di-jet invariant mass
at a similar rate as the ratio of the $\ttbar\Pg\Pg$ to $\ttbb$ cross sections.
For instance, using LO matrix elements,
we checked that both quantities increase by a factor two in the range between 100 and 250\,GeV.

The plots in \reffi{fig:ttbb100}, where an additional cut
$m_{\Pb\Pb}>100~\GeV$ is applied, reveal distinctive kinematic features 
of the MC@NLO enhancement in the Higgs-signal region.  The
unambiguous MC@NLO/NLO peaks that appear in the distributions, both
in the transverse momentum of the first b-jet (\reffi{fig:ttbb100}.a) and in the $\Delta R$
separation of the first two b-jets (\reffi{fig:ttbb100}.b), show
that the MC@NLO enhancement is dominated by back-to-back b-jets with the
smallest possible $p_\rT$ that is needed to reach $m_{\Pb\Pb}=100~\GeV$.
This is consistent with the expected behaviour of 
double $\Pg\to\bbbar$ splitting contributions in \reffi{fig:topologies}.b,
where emissions at small-$p_\rT$ are doubly enhanced by 
soft and collinear singularities associated with the 
parent gluons.
Also this interpretation is  {fully confirmed} by the fact
that MC@NLO-induced shape distortions in \reffi{fig:ttbb100}
disappear {almost completely} 
when $\Pg\to\bbbar$ shower splittings are switched off.

To exclude the possibility that 
double splittings in our simulation are artificially enhanced by a
too high choice of the resummation scale, we checked
that the characteristic ``double-splitting'' enhancement in the
$m_{\bbbar}$ distribution of Fig.~\ref{fig:ttbb} is present also 
in simulations based on merged LO matrix elements for 
$\ttbar$ plus multi-jet production. 
In this framework, $\ttbb$ events are not showered 
with a global resummation scale, but 
starting from a scale that is determined 
according to the most likely shower history of the
event at hand.
Comparing the shape of the MC@NLO distribution of Fig.~\ref{fig:ttbb}
against MEPS@LO simulations~\cite{Hoeche:2009rj} of $\ttbar+\le 3j$
with massive b-quarks, we found good agreement for 
merging scales around $15~\GeV$, \ie for the case where 
most of the phase space associated with (the first) $\Pg\to\bbbar$ 
splittings is described in terms of matrix elements, 
as in the present MC@NLO simulation. 
A thorough understanding of the uncertainties related to the choice of the
merging scale and the interplay between matrix elements and 
parton shower in the vicinity of the kinematic threshold
for $\Pg\to\bbbar$ splittings requires further detailed studies
that are beyond the scope of this letter.

In summary, we presented the first complete MC@NLO simulation 
of $\ttbb$ production at the LHC, including 
b-quark mass effects.
This allows one to cover the full $\ttbb$ phase space 
at NLO accuracy and to describe contributions stemming from double
collinear $\Pg\to\bbbar$ splittings, which can lead to a significant
contamination of the $\ttbar\PH(\bbbar)$ signal.
This unexpected finding changes the standard perturbative picture of 
$\ttbb$ production based on hard b-quark jets.
The presented simulation
will allow for a thorough analysis of the related uncertainties.  In this
respect it will be important to assess the role of the parton-shower tune and to devise
efficient strategies for the rejection of double-splitting contributions. 
Aspects not discussed here, such as top-quark decays, hadronisation and underlying events,
can be simulated in a fully automated way using \Sherpa.  
To gain more insights into theoretical uncertainties 
associated with the parton shower and the b-quark mass, it will be very instructive 
to compare the four-flavour
scheme adopted in this paper to the five-flavour scheme.
Both schemes provide reliable NLO predictions for observables 
involving resolved b-jets at the LHC~\cite{Maltoni:2012pa}.
In the five-flavour scheme, where b-quarks are massless,
$\ttbb$ matrix elements cannot be used to fill the entire b-quark phase space,
and the collinear regions need to be described
by lower-multiplicity hard matrix elements ($\ttbar\Pg$, $\ttb$, $\ttbar$, etc.)
supplemented by parton-shower emissions. Technically this requires the merging of 
NLO matrix elements for $\ttbar+0,1,2$ jets, which was presented for the first time 
in~\cite{Hoeche:2014qda}. A consistent combination of this recent simulation 
and the massive $\ttbb$ predictions presented in this paper would
provide an optimal description of $\ttbar$ plus multi-jet production.

\section*{Acknowledgments}

We thank A.~Denner, S.~Dittmaier and L.~Hofer for providing us with the
one-loop tensor-integral library \Collier.  We are grateful to
S.~H\"oche, F.~Krauss  and M.~Sch\"onherr for several fruitful
discussions and technical support. 
We thank 
Graeme Watt for generating an MSTW set of 
4F PDFs with 5F $\alphaS$ running.
The research of F.~C., P.~M., N.~M.~and S.~P~is funded by the SNSF 
and supported, in part, by the European
Commission through the 
network PITN-GA-2010-264564.
The work of F.~S.~was supported by the German Research Foundation 
via grant
DI 784/2-1.


\end{document}